# GaVe: A Webcam-Based Gaze Vending Interface Using One-Point Calibration

Zhe Zeng
Technical University of Berlin, Germany

Sai Liu
Technical University of Berlin, Germany

Hao Cheng
University of Twente, Enschede, Netherland

Hailong Liu
Nara Institute of Science and Technology, Japan

Yang Li
Karlsruhe Institute of Technology, Germany

Yu Feng*
Leibniz University Hannover, Germany

Felix Wilhelm Siebert
Technical University of Denmark, Lyngby, Denmark

*Corresponding Author

Gaze input, i.e., information input via eye of users, represents a promising method for contact-free interaction in human-machine systems. In this paper, we present the GazeVending interface (GaVe), which lets users control actions on a display with their eyes. The interface works on a regular webcam, available on most of today's laptops, and only requires a short one-point calibration before use. GaVe is designed in a hierarchical structure, presenting broad item cluster to users first and subsequently guiding them through another selection round, which allows the presentation of a large number of items. Cluster/item selection in GaVe is based on the dwell time, i.e., the time duration that users look at a given Cluster/item. A user study (N=22) was conducted to test optimal dwell time thresholds and comfortable human-to-display distances. Users' perception of the system, as well as error rates and task completion time were registered. We found that all participants were able to quickly understand and know how to interact with the interface, and showed good performance, selecting a target item within a group of 12 items in 6.76 seconds on average. We provide design guidelines for GaVe and discuss the potentials of the system.

## Introduction

Since touchless input is not only convenient but also hygienic, the Covid-19 pandemic has led to a rise in demand for touchless human-machine interaction in the public space. Especially in high-traffic fast-food restaurants and public transportation ticket offices, touchless ordering and ticketing systems are needed to prevent the transmission of viruses. Touchless gaze-based input represents a promising method for interaction in touchless human-machine interaction (HMI) systems.

In daily life, humans use their eyes mainly to obtain information, but methods have been also developed to use eye as an input modality in HMI. For example, various interfaces have been developed which let users control websites (Menges, Kumar, Müller, & Sengupta, 2017), enter text (Feng, Zou, Kurauchi, Morimoto, & Betke, 2021; Lutz, Venjakob, & Ruff, 2015; Majaranta & Räihä, 2002) or PIN codes (Best & Duchowski, 2016; Cymek et al., 2014) with their eyes. Poitschke, Laquai, Stamboliev, and Rigoll (2011) demonstrated that gaze-based interaction can be superior over conventional touch interfaces in the automotive environment. For public displays, gaze-input has multiple advantages. First, gaze input facilitates touchless interaction with the interface, which prevents the transmission of e.g., viruses through touch between multiple users. Second, gaze input can prevent shoulder surfing and ensures user privacy when using public displays. Third, as the price of commercial eye tracker devices is decreasing, it presents a cost-efficient input method. More recently, gaze estimation has been conducted on off-the-shelf consumer hardware such as webcams (Liu, Lee, Rajan, Sluzek, & McKeown, 2019; X. Zhang, Sugano, & Bulling, 2019). This makes gaze estimation technically and economically feasible for all devices that include a front camera, such as cellphones, tablets, and laptops. Thus, using gaze input is no longer limited by high hardware costs and can be used to benefit a much larger user group.

Despite these advances, gaze-based interaction is still facing a number of challenges that need to be considered in the design of interfaces:

1) The "Midas touch" problem (Jacob, 1990)—Searching and selecting an interactive item are not always clearly separated. It can be challenging to distinguish a user just looking at an object on a screen from the intention of the user to select that object.

2) The calibration requirement—The process is considered time-consuming. Users need to re-calibrate multiple times per day to ensure the eye tracking quality (Feit et al., 2017). To enable gaze interaction on a public display, the system should attempt to avoid or shorten the calibration process to improve user acceptance and experience.

3) The noise come from user—Noise in eye tracking data always accompanies in gaze-based interaction, such as head movement (Kowler, 2011) and drift (Robinson, 1968). Those noise affect the accuracy and precision of the eye tracking data.

To address the aforementioned challenges, we propose a novel gaze interface, which can be built using an off-the-shelf webcam. The "Midas touch" and noise problems are addressed through high spatial separation of interactive display elements, while the calibration requirement is achieved through a brief one-point calibration. The contributions of this work are as follows:

1) We present a touchless gaze input method with low spatial accuracy for eye tracking data, i.e., without personal calibration, using a single off-the-shelf camera;

2) We develop a vending-machine gaze interface prototype (GaVe) (as shown in Figure 1), where the visual search area and the interactive buttons are spatially separated, i.e., content items are displayed in the center of the screen, with interactive buttons placed at the edges to reduce eye movement and head movement during visual search;

3) We conduct a user study with the functional GaVe interface and evaluate the usability of the interface. Relevant parameters, i.e., size of the central visual search area, distance from a user to the screen, and the duration of dwell time, are compared and analyzed. The main findings are summarized in a design guideline.

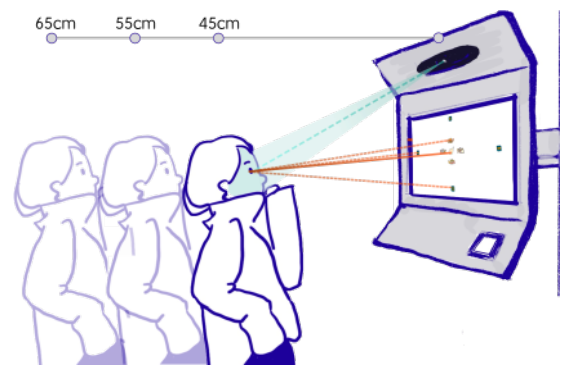

Figure 1. Visualization of ordering on a vending machine with gaze interaction in the real-world application.

## Related work

### Gaze interaction

Human gaze can contain complex information about a person's interests, hobbies, and intentions (Kowler, 2011). To leverage this information, eye tracking technologies are applied to measure eye positions and movements. They have been widely used in medical, marketing, and psychological research. Moreover, with the help of eye tracking, eye has been transformed into an alternative input modality for controlling or interacting with other digital devices (Jacob, 1990). Today, there are multiple, functionally different ways to use eye as an input modality. The most popular gaze-only designs are summarized in the following paragraphs.

**Dwell-based gaze interaction** People's perception of a stable visual information is achieved by fixation (Martinez-Conde, Macknik, & Hubel, 2004). In dwell-based gaze interaction, fixation duration, i.e., dwell time, is used to activate an action and the eye position is used to replace the mouse cursor on the screen. Dwell-based gaze interaction is subject to the "Midas touch" problem, i.e., a difficulty to distinguish between information intake and object selection on a display. To address the "Midas touch" problem, a time-based threshold is set for the selection of an object. Only if this pre-defined dwell-time threshold has been reached, will the corresponding action be triggered. Generally, the setting for the dwell-time threshold varies from 200ms to 1000ms (Majaranta, Ahola, & Spakov, 2009; Majaranta & Räihä, 2002; Møllenbach, Hansen, & Lillholm, 2013). Hence, prior to the implementation of interfaces, an assessment of dwell time thresholds for a specific task can be necessary. Thanks to the straightforward function and its easy implementation, dwell-based gaze interaction has become one of the most popular gaze interaction methods (Majaranta, Ahola, & Spakov, 2009; Majaranta & Räihä, 2002; Mott, Williams, Wobbrock, & Morris, 2017). High tracking accuracy emerges as an additional challenge for dwell-based gaze interaction systems, as the method relies on a relatively high accuracy to correctly register the spatial location of the fixated object on the display. Hence, a calibration of the tracking system is needed and the size and spatial separation of the interactive items in the display can influence detection performance (Ware & Mikaelian, 1986).

**Blink-based gaze interaction** In blink-based systems, the action of closing ones' eyes is used to trigger an action in the interface. To prevent unintentional triggering of actions through involuntary blinks, only voluntary blinks are used for gaze interaction. Frequently, voluntary blinking is defined over blink-duration (Grauman, Betke, Lombardi, Gips, & Bradski, 2003), with blinks over 200ms registered as voluntary (Ashtiani & MacKenzie, 2010) or using single eye closure as a trigger method (Ramirez Gomez, Clarke, Sidenmark, & Gellersen, 2021). Similar to the dwell-based gaze interaction method, eye position is used to control the cursor on the device's screen and its performance is similarly influenced by the accuracy of eye tracking.

**Gesture-based gaze interaction** Differing from the above two methods, gesture-based gaze interaction utilizes intentional saccades to trigger actions on a display. Saccades occur when the human gaze voluntarily or reflexively "jumps" from a fixated point to a desired end point (Duchowski, 2017). Eye gestures are defined as an ordered sequence of intentional saccades (Drewes & Schmidt, 2007). They consist of different "paths" of saccades which can be mapped to specific interaction commands. Eye gesture-based interaction has several advantages over dwell- and blink-based gaze interaction. Firstly, eye gestures can distinguish intentional interaction commands from unintentional commands, thus effectively solving the "Midas touch" problem. Secondly, compared to dwell-based interaction, the control area of eye gestures does not rely on the exact position of gaze data, just on the relative position between starting and end points of saccades. However, there is a considerable disadvantage of gesture-based methods. Users of this interaction method need to learn and remember the defined gaze gestures before using them. This heavily limits its applicability in public displays.

**Pursuit-based gaze interaction** Smooth pursuit eye movements occur when the eyes follow a moving object. Pursuit-based interaction is established by matching the trajectories of eye-movement to moving object trajectories on a display (Vidal, Bulling, & Gellersen, 2013). Different types of trajectories can be used, e.g., circular trajectories (Esteves, Velloso, Bulling, & Gellersen, 2015; Niu et al., 2021), linear trajectories (Porta, Dondi, Pianetta, & Cantoni, 2021; Zeng, Neuer, Roetting, & Siebert, 2022; Zeng & Roetting, 2018; Zeng, Siebert, Venjakob, & Roetting, 2020), and irregular trajectories such as an object's outline (Sidenmark, Clarke, Zhang, Phu, & Gellersen, 2020). In comparison to the other gaze interaction methods mentioned above, pursuit-based gaze interaction does not require precise gaze coordinates or personal calibration for a robust gaze-based interaction. As a dynamic interface, pursuit-based interaction is much different from existing human-machine interfaces. We need to consider the user acceptance when designing pursuit-based interfaces.

### Gaze estimation using a webcam

Eye tracking relies on technology that can register eye position and eye movement. Most eye tracking devices combine a camera and infrared light (IR) sources to estimate the gaze position, using the IR light to position the eyes in relation to the camera.

Recently, off-the-shelf cameras have been used to estimate eye positions (D. Hansen, Hansen, Nielsen, Johansen, & Stegmann, 2002; Papoutsaki, Laskey, & Huang, 2017). Some studies developed interaction systems using gaze direction detection to enter text (C. Zhang, Yao, & Cai, 2018; X. Zhang, Kulkarni, & Morris, 2017) or PIN (Khamis, Hassib, Zezschwitz, Bulling, & Alt, 2017). However, while results are promising, the spatial accuracy of off-the-shelf camera gaze estimation is still relatively low. The gaze estimation error is around 5-6° for model-based methods and 2-4° visual angle for appearance-based estimation methods (X. Zhang et al., 2019). To circumvent the low spatial accuracy problem, Hansen et al. (2002) propose to utilize large interactive items.

In addition to the accuracy problem, the time needed for calibration may affect a user's acceptance and experience. This motivates researchers to design applications that work without personal calibration, such as using smooth-pursuit movements based on the front-facing camera of a tablet Eyetell (Bafna, Bækgaard, & Paulin Hansen, 2021). This calibration-free design is appealing when it comes to the use of public displays. Thus, to facilitate the implementation and public displays of vending ordering, in this work, we focus on developing a dwell-based gaze interface without a lengthy calibration process using an off-the-shelf camera.

## Methods

Weighing the advantages and disadvantages of available gaze-based interaction systems, we implemented a dwell-time based gaze interaction system with a brief (2-second) one-point calibration. It is characterized by ease of understanding and implementation. Our system is implemented on an off-the-shelf webcam and uses facial landmarks and a shape-based method to estimate the direction of gaze.

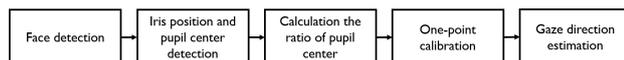

Figure 2. The image processing pipeline for gaze estimation in GaVe.

As shown in Figure 2, the gaze estimation module consists of the following parts: (1) face detection, (2) iris position and pupil center detection, (3) estimating the ratio of each pupil center, (4) one-point calibration, and (5) the five gaze directions estimation, i.e., right, left, up, down, and center. Steps 1 and 2 use the 68-point face detection method implemented by the open-source Python Dlib library (Lamé, 2019), resulting in an initial rough estimation of a user's pupil center. In steps 3-5, we optimize the gaze estimation to detect five gaze directions using one-point calibration. In the following, we explain our method in detail.

**(1) Face detection** The Dlib's 68-point facial landmark is used to detect a frontal face. In the face detection, 12 points are used for detecting eyes (6 points for each eye). As shown in Figure 3, the point landmarks 36-41 are for detecting the left eye and the point landmarks 42-47 are for detecting the right eye.

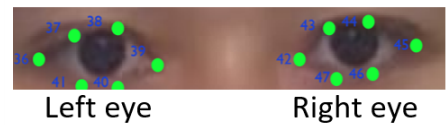

Figure 3. An example of eye-landmarks detected using the Python Dlib library.

**(2) Iris position and pupil center detection** After having identified the areas containing the eyes using the Dlib's 68-point facial landmark, we further partitioned the eye image into the left-eye and right-eye images. The two images were then analyzed individually for detecting their corresponding iris positions. There is a sharp boundary between the sclera and the iris, so the corresponding limbus can be easily obtained by image processing (Holmqvist et al., 2011). A bilateral filter is used to filter and erode the image in order to smooth it and enhance the color of the iris (Tomasi, & Manduchi, 1998). Both eye images are converted into a binary mask—the iris contours are denoted in black color, in order to distinguish the iris from the other parts of the eye. The center coordinates of the pupil for each eye are finally derived as the centroid of the iris contour by calculating the image moments.

**(3) Calculating the ratio of pupil center** The ratio of the pupil center is calculated based on its center position in relation to the edge positions that the pupil can normally reach. The following formula denotes the calculation of the horizontal ratio of the left pupil:

$$h_{ratio\_left} = \frac{x - x_{min}}{x_{max} - x_{min}} \quad (1)$$

where $x$ is the center $x$-coordinate of the left pupil extracted from the above steps and $x_{max}$ and $x_{min}$ are the maximum and minimum values of the eyelid edge, that the pupil can reach. The value of $h_{ratio\_left}$ is the horizontal ratio for left eye which ranges from 0 to 1. When the ratio is close to 1, it means that the participant is looking in the leftmost direction. When the ratio is close to 0, it means that the participant is looking in the rightmost direction. When the ratio is close to 1, it means that the participant is looking in the leftmost direction. When the ratio is close to 0, it means that the participant is looking in the rightmost direction. The points 36 and 39 correspond to the corners of the left eye. Here, the x coordinate of point 36 is used for $x_{min}$ and the x coordinate of point 39 is $x_{max}$.

Based on the observation that the pupil rarely reaches the eyelid edge denoted by the landmark positions, e.g., $p_{36}$ and $p_{39}$, we optimized the maximum and minimum values using a pilot study. Seven participants (4 males and 3 females) were asked to record their pupil movement by orienting to the eyelid edge. They were asked to keep their head still while facing a display screen in front of them, and then to look left, right, up and down as far as they would. We recorded the data for each direction and calculated the average $h_{ratio}$ and $v_{ratio}$. The mean of the $h_{ratio}$ is 0.28 when participants look rightmost and 0.87 when they look leftmost. These two values are used as $h_{ratio\_min}$ and $h_{ratio\_max}$. The vertical ratio is calculated in the similar way as the horizontal. In the vertical direction, the points 37 and 38, 40 and 41 refer to the upper and lower eyelids of the left eye, respectively, where $y_{min} = (y_{37} + y_{38})/2$ and $y_{max} = (y_{40} + y_{41})/2$. The average ratios are 0.48 when gazing at the top and 0.95 when gazing at the bottom. These two values are used as $v_{ratio\_min}$ and $v_{ratio\_max}$. Thus, we re-normalize the ratios using the data from pilot study (see formula 2 for left pupil).

$$h_{ratio\_left\_optimized} = \frac{h_{ratio\_left} - h_{ratio\_min}}{h_{ratio\_max} - h_{ratio\_min}} \quad (2)$$

The final horizontal ratio of the pupil center is the averaged value of both the left and right eyes. It should be noted that the right eye is estimated using the same method.

$$h_{ratio\_final} = \frac{h_{ratio\_left\_optimized} - h_{ratio\_right\_optimized}}{2} \quad (3)$$

With the optimized minimum and maximum ratios acquired from the pilot study, we were able to extend the original gaze tracking method for the vertical direction.

**(4) One-point calibration** Calibration is the process of mapping the local eye coordinates obtained from the eye-tracker/camera to a specific point on the display (resolution 1920 × 1080 pixels). For GaVe, we use a one-point calibration to simplify the process, and ensure a short calibration time during walk-up-and-use scenarios. The calibration is visualized in Figure 4. At the start of the calibration, a red point is displayed in the center of the screen. Since the blink rate is 17 blinks/min at rest (Bentivoglio et al., 1997), that is, on average, people blink once every 3-4 seconds, so we took two seconds to ensure both the quality of the collected data and comfortability. After two seconds, the point turns green. The participants are instructed to keep their heads still and look at the red point until it turns green. Figure 4 shows the process of calibration, the coordinates of the calibration point, $x_{screen}$, $y_{screen}$, are (960, 540) on the screen. For example, one participant's data produces a detected horizontal and vertical ratio, denoted as $h_c$, $v_c$, of (0.56, 0.51), respectively. These individual $h_c$, $v_c$-ratios are set as the central point of the screen for the individual user. It should be noted that the $h_c$, $v_c$ ratios vary slightly across users.

**(5) Gaze direction estimation** The one-point calibration results in the horizontal and vertical ratio of the central point $h_c$, $v_c$. According to the data from our pilot study, the individual ratio of the central point can vary slightly around the actual central point of the screen.

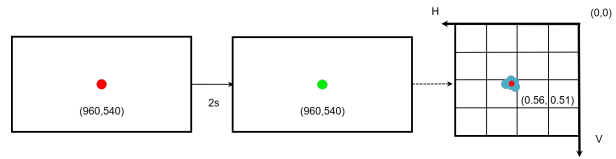

Figure 4. Illustration of the one-point calibration process for one participant.

In the pilot study mentioned in step (3), after completing the first task, i.e., looking at four directions as far as possible, all the participants completed another task of looking at the four targets on the screen (top, bottom, left, and right) and midpoint for one-point calibration in turn, the ratios were recorded. We found a central space, more precisely, it is a rectangle-like space (see Figure 5).

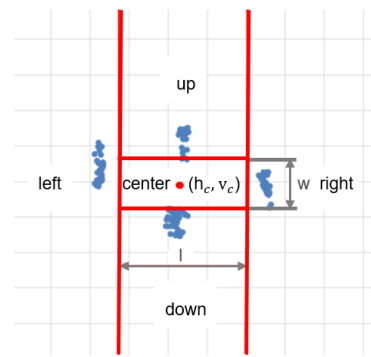

Figure 5. Visualization of the different functional zones on the GaVe display for one participant's data during the pilot study. The blue points are the ratios recorded for four targets, and the red point is the horizontal and vertical ratio of the central point $h_c$, $v_c$ after one-point calibration.

The width and length of the rectangle are denoted as *w* and *l*. Based on the pilot study, we calculated the approximate ratios of *w* and *l* in relation to the actual width and height of the screen. We found that 0.4 and 0.2 are the proper values to suit all participants. Therefore, we adopted these values for our final mapping from the gaze position to the screen position. More specifically, the total area of the screen was partitioned into the center (the rectangular region), left, right, up, and down. When the gaze was mapped in the corresponding area on the target screen, a gaze event, i.e., "look right", "look left", "look up", "look down" and "look center" would be detected.

### Interface

The stages of the selection process of GaVe are visualized in Figure 6. All menu items are located in the center of the screen. As shown in Figure 6 (a), there are four clusters in the initial interface, arranged in the four directions up, down, left, and right. In each of the four clusters, three items are grouped, e.g., the top cluster combines a pizza, a burger, and a hot dog. When the system detects that the user is looking towards the center of the interface, i.e., the inactive area within the rectangle defined in Figure 5, no action is triggered and the interface shows the cluster selection screen (Figure 6(a)). GaVe stays this initial interface, as long as no looking up, down, left, or right is detected

Four arrows are located outside of the central area. Once a user's gaze is detected in one of the four interactive directions (defined in Figure 5) the corresponding arrow is marked with a gray circle, as real-time visual feedback. If the user continuously focuses on an arrow for longer than a predefined time threshold, the circle around the arrow turns red to confirm the selection.

The system has a two-stage project selection process: Cluster and item selection. For example, the target item ("chicken drumstick") is presented in the middle of the screen in Figure 6.

**Cluster selection** The first stage of cluster selection is illustrated in Figure 6(a-c). In Figure 6 (a), the user selects the lower cluster consisting of "chicken drumstick-chips-popcorn" by looking at the down arrow. As shown in Figure 6 (b), the down button is highlighted with a gray circle to show that this button is in focus. As shown in Figure 6 (c), if the user continuously looks at this button for a predefined time threshold, the circle turns red to confirm the first stage of the cluster selection.

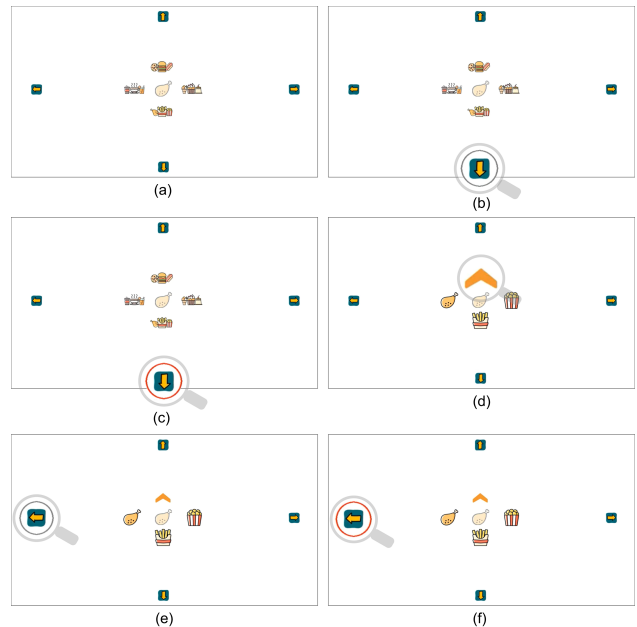

Figure 6. The two-stage item-selection process of choosing an item in GaVe. The user first selects a target cluster and then selects the desired item from the cluster. The gray circle in the figure is the real-time feedback on which item the user is looking at, and the red circle is the feedback about the confirmation of a selection. The semi-transparent item is the given target. The gray magnifying glass marks the detected eye position, which does not appear in real interactions.

**Item selection** The second stage of item selection is illustrated in Figure 6(d-f). In this stage, the items in the selected cluster are expanded, as shown in Figure 6 (d). To help the user keep track of items in each cluster, the item located on the right side of the original cluster is also displayed on the right side in this stage. The same goes for the item originally located on the left side of the cluster, it is moved to the left side in the item-selection stage. The item originally located in the middle of the cluster is moved down to the bottom position. In the top position, a back button appears, allowing users to go back to the cluster stage. In Figure 6 (e), the target "chicken drumstick" is located on the left. As the selection continues, the user needs to look at the left side of the interface. Again, the gray circle gives feedback to the user that the left side of the interface is detected. As the final step, Figure 6 (f) shows the confirmation interface when the target item ("chicken drumstick") is selected. If a time of 10 seconds in the selection process of one interface stage is exceeded, the system will reset to the initial cluster-selection screen, and jump to the next target item. The previous item-selection task is then registered as a missed selection.

# User study

To explore the usability of the GaVe interface, we implemented the interface in a stylized vending machine, using a webcam to register participants' gaze. In the study, we experimentally varied the threshold for dwell time that triggers an action, the distance from the user to the screen, and size of the central area for the vending machine to identify the optimal setup for the interface.

## Participants

In total 22 participants were recruited for the experiment (13 males, 9 females, mean age: 28.1 years, ranging from 23 to 40 years). Ten of they wore glasses and 3 participants wore contact lenses. The remaining 9 participants did not wear visual aids. Most of the participants had no experience with eye tracking and gaze interaction.

## Apparatus

A 15.6" laptop with Intel Core i5-7300HQ 2.8GHz and 8GB RAM was used for the registration of participants gaze and for displaying the GaVe interface on its 1920×1080 pixels screen. The embedded webcam has a resolution of 1080p. At a distance of 45cm from the participant to the screen, 44 pixels correspond to 1° visual angle. An external light source (20W Halogen Lamp) was set directly behind the webcam, as shown in the Figure 7 to ensure adequate ambient lighting.

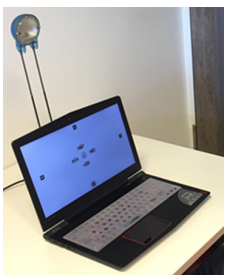

Figure 7. The experiment setup.

## Experiment design

The study used a three-factorial within-subjects design. The independent variables are:

- Size of central area (small, medium, large)
- Distance from user to screen (45, 55, 65cm)
- Dwell time threshold (0.5, 0.8, 1.0, 1.2s)

The pilot study suggested that the ratio of the central area was 0.4 in the horizontal and 0.2 in the vertical directions. Thus, the central area measured by the ratio relative to the screen size in horizontal and vertical directions is set to 0.16×0.09 (small), 0.2×0.12 (medium), and 0.24×0.16 (large), respectively. Estimated at 45 cm from the screen, the range of visual angle in horizontal and vertical directions is set to 10.72°×7.74°, 13.4°×10.32°, and 16.08°×13.76° corresponding to the above central area in each size. In total, there are 36 (3×3×4) different combined conditions, and each condition was repeated 4 times for each participant. This yielded a total number of 3168 trials.

The participants' performance was assessed through objective and subjective criteria. The objective criteria included task completion time and error rate. The task completion time is defined as the time that participants take to complete a trial, i.e., to finish the selection of a given target item. Errors are registered when participants select an item that is not the current target item, or they are unable to interact with the system for a predefined time, i.e., no action (cluster/item selection) is registered after 10s. The error rate is calculated as the fraction of the number of trials registered as errors divided by the total number of trials.

The subjective experience of participants was assessed after completion of the experiment. The participants were asked three questions: 1) Which is the most comfortable distance for you? 2) At what distance do you think you can select the target most accurately? 3) How would you evaluate this gaze interaction system?

## Procedure

The experiment was conducted in eye tracking laboratory of the Chair of Human-Machine Systems at the Technical University of Berlin. Before the experiment, all participants signed an *informed consent form* and answered a demographic questionnaire. After this, the participants were given a short introduction to the system and received an explanation on how to use it. The participants were instructed to select a given target item as accurately and quickly as possible. The given target item was displayed in the central area of the interface in a semi-transparent form. Each experiment condition included four trials, i.e., repeated four times. All participants performed the experiment in a seated position to adjust and stabilize the interaction distance. The order of the distance conditions was controlled among participants to avoid a frequent change of the seated position. Half of the participants were tested in the order from far to near, and the other half in the order from near to far. Under each distance, the orders of the conditions in terms of the central area size and dwell time were randomized across participants to prevent the occurrence of effects of sequence. After completing the task for all conditions, the participants were asked to answer the three open questions listed above. During the experiment, the participants were allowed to rest when one condition was finished. The experiment lasted approximately 30 minutes.

Table 1. Mean and standard deviation for task completion time and error rate across different experimental conditions (user to screen distance; dwell time; central area size).

| Distance (cm) | Dwell time (s) | Task completion time | | | | | | Error rate | | | | | |
|---|---|---|---|---|---|---|---|---|---|---|---|---|---|
| | | Large | | Medium | | Small | | Large | | Medium | | Small | |
| | | M | SD | M | SD | M | SD | M | SD | M | SD | M | SD |
| 45 | 0.5 | 4.55 | 1.17 | 4.21 | 1.28 | 3.86 | 1.20 | 0.32 | 0.28 | 0.35 | 0.26 | 0.43 | 0.33 |
| | 0.8 | 5.38 | 1.44 | 6.11 | 1.77 | 5.82 | 2.28 | 0.26 | 0.31 | 0.16 | 0.25 | 0.24 | 0.25 |
| | 1.0 | 7.32 | 1.44 | **6.90** | **1.61** | 6.95 | 1.74 | 0.18 | 0.25 | **0.07** | **0.22** | 0.16 | 0.29 |
| | 1.2 | 9.21 | 1.91 | 8.66 | 2.03 | 8.59 | 1.93 | 0.27 | 0.33 | 0.16 | 0.27 | 0.24 | 0.25 |
| 55 | 0.5 | 4.45 | 1.55 | 4.51 | 2.51 | 4.34 | 1.66 | 0.40 | 0.38 | 0.43 | 0.39 | 0.40 | 0.31 |
| | 0.8 | 6.23 | 1.82 | 5.69 | 1.68 | 5.92 | 2.04 | 0.20 | 0.27 | 0.33 | 0.29 | 0.49 | 0.34 |
| | 1.0 | 8.21 | 2.34 | 7.24 | 1.71 | 7.44 | 2.40 | 0.34 | 0.33 | 0.16 | 0.24 | 0.15 | 0.31 |
| | 1.2 | 9.21 | 2.28 | 9.72 | 2.51 | 8.91 | 1.86 | 0.39 | 0.32 | 0.33 | 0.34 | 0.38 | 0.33 |
| 65 | 0.5 | 4.55 | 2.30 | 4.78 | 2.71 | 5.27 | 1.69 | 0.61 | 0.38 | 0.59 | 0.33 | 0.56 | 0.42 |
| | 0.8 | 5.95 | 2.60 | 5.42 | 2.13 | 5.34 | 1.81 | 0.41 | 0.36 | 0.45 | 0.41 | 0.41 | 0.38 |
| | 1.0 | 8.33 | 2.30 | 8.34 | 2.31 | 7.76 | 1.96 | 0.49 | 0.35 | 0.39 | 0.32 | 0.43 | 0.36 |
| | 1.2 | 8.87 | 2.00 | 9.40 | 2.21 | 9.87 | 2.62 | 0.56 | 0.34 | 0.43 | 0.32 | 0.48 | 0.37 |

## Results

In the following, the objective measures of the user study will be presented, followed by the subjective assessment by the participants. A three-way repeated-measures ANOVA (3*3*4) was conducted for the data analysis. The Shapiro-Wilk test and Q-Q-Plot were used to validate the assumption of data normality. We used the Greenhouse–Geisser correction when the Mauchly's sphericity test indicates that the data does not fulfill the sphericity assumption. Moreover, Bonferroni correction was applied for post-hoc pairwise comparison.

For *task completion time* and *error rate*, detailed results are presented in Table 1 for all experimental conditions. A detailed analysis of this finding is given in the following subsections. It can be observed that at the distance 45cm from the user to the screen with the one-second dwell time and the medium-sized screen central area, the participants achieved the minimum error rate with a relatively short task completion time (highlighted in boldface with an underline).

## Task completion time

Figure 8 visualizes the task completion time for different distances to the screen and different dwell time conditions.

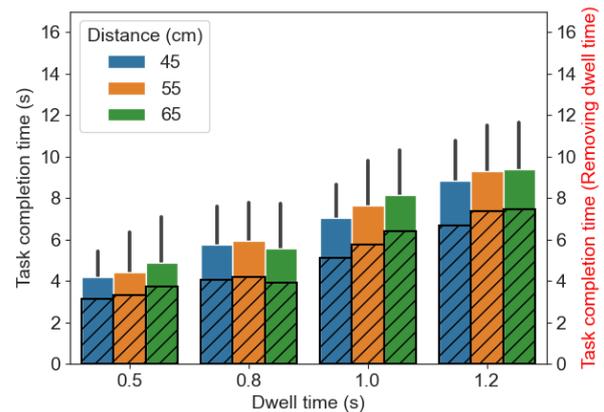

Figure 8. The average task completion time. The error bars represent the standard deviation in each condition. The striped bars are the task completion time after removing the duration of the dwell time.

It can be observed that task completion time is closely associated with the dwell time—the task completion time increased alongside the dwell time set in the conditions. To further analyze the impact of the task completion time under different distance conditions, we removed the fixed duration of the dwell time. The results are illustrated by the black striped bars within the original bar plots in Figure 8. Even after removing the fixed duration of the dwell time for activating an action, we still found that the task completion time is longer for the dwell time conditions of 1.0 and 1.2s than that of 0.5 and 0.8s.

There was a significant main effect of dwell time ($F(2.05, 43.14) = 153.15, p < .001$). The pairwise comparisons show that all comparisons between conditions were significant, i.e., between 0.5&0.8s, 0.5&1.0s, 0.5&1.2s, 0.8&1.0s, 0.8&1.2s, and 1.0&1.2s. No significant main effect was found for the size of the central area ($F(2, 42) = 0.81, p = .45$) and distance from the user to the screen ($F(2, 42) = 3.16, p = .05$). Furthermore, we found no two-way and three-way interaction of factors.

### Error rate

Since the Shapiro-Wilk test shows that the error rate is not normally distributed ($p < .05$), we applied an Align Rank Transform (Wobbrock, Findlater, Gergle, & Higgins, 2011) before the repeated ANOVA.

As shown in Figure 9(a), the error rate decreased when the dwell time increased from 0.5 to 1.0s, and reached its lowest at 1.0s. The error rate rose once again when the dwell time was longer (1.2s). The error rate is lower for shorter distances at all dwell time levels. In terms of error rate, the dwell time ($F(3, 735) = 15.02, p < .001$) and distance from the user to screen ($F(2, 735) = 48.75, p < .001$) had a significant effect. The difference is significant between 0.5&0.8s ($p < .001$), 0.5&1.0s ($p < .001$), 0.5&1.2s ($p < .01$), and 1.0&1.2s ($p < .001$); and significant differences in terms of the distance were found between 45&55cm, 45&65 and 55&65 ($p < .001$). However, there was no significant difference regarding the size of the central area ($F(2, 735) = 1.77, p = 0.17$).

We found an interaction effect between the dwell time and the size of the central area ($p < .05$) with respect to the error rate. Namely, the error rate at the 0.5s dwell time is significantly higher than that at the 1.0s dwell time for both the small-sized area condition ($p < .01$) and the medium-sized area condition ($p < .001$).

We further analyzed the error rate by distinguishing between missed detections (Figure 9(b)) and false detections (Figure 9(c)). A false detection was registered when the selected item is not the given target item. A missed detection is registered when a participant does not activate an action within the predefined time frame of 10s. Figure 9(b) and (c), visualize how the false detection rate—the fraction of the false scenarios over the total number of trials—*decreases* gradually from 0.5 to 1.2s of the dwell time, while the missed detection rate *increases* gradually from 0.5 to 1.2s.

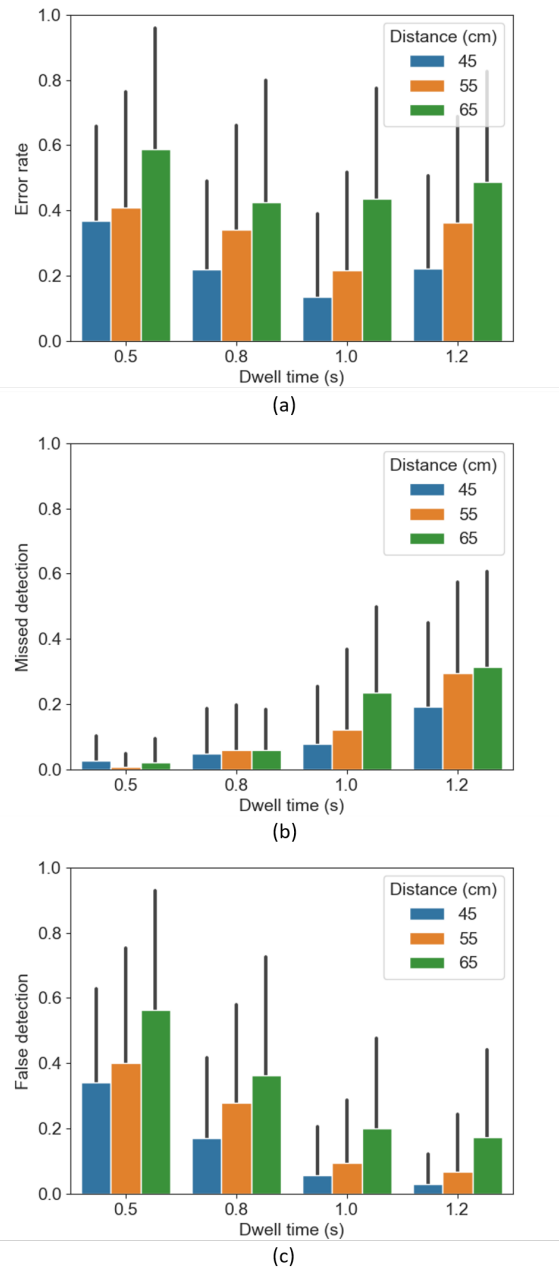

Figure 9. The average error rate, false detection rate and missed detection rate, error bars represent the standard deviations.

### Subjective evaluation

Besides the objective variables, i.e., task completion time and error rate, we also collected subjective feedback from the participants. In terms of the comfortable distance to the screen, 55% of the participants thought that the smallest distance of 45cm was the most comfortable condition to accomplish the given tasks, while 32% of the participants considered that the most comfortable distance was 55cm, and only the remaining 13% chose 65cm. In terms of system accuracy, 91% of participants felt that the system was most accurate at 45cm, while only 9% of the participants preferred the distance of 55cm, and no participant perceived the largest distance of 65cm as the most accurate one. Asked about their general evaluation of the system, many participants considered that this gaze interface was innovative. One participant from the medical specialty mentioned that this touchless interaction was hygienic, and gave the interactive system a highly complimentary remark.

# Discussion

The need for touchless input modality is particularly increasing during Covid-19 pandemic. The aim of this study was to design a webcam-based gaze interface for touchless human-computer interaction on public displays. We developed a gaze-based interface for a vending machine, in which the interaction is triggered by the gaze direction estimation registered through a webcam. User can complete an input with a short calibration at the low spatial accuracy of eye tracking. Compared to traditional eye-tracking devices, our method has a lower device cost.

A controlled laboratory experiment was conducted to study the usability of the system and to comprehensively assess optimal system parameters. From the user study, we found that the GaVe interface is effective and easy to use for most participants, even for participants wearing contact lenses and glasses. All participants were able to use the system after a short introduction.

The result of user study showed that there was a marked increase in the task completion time when the dwell time became longer, even when accounting for the longer wait times during the dwell-time based trigger. One possible reason for this result is that the excessive duration of dwell time strains the eyes, which in turn increases the difficulty of the dwell-based selection (Majaranta, Ahola, & Spakov, 2009). There were more selections that had been interrupted by a failure to maintain a fixation on the target for the required dwell time. The user needs to try for a longer time to successfully select the target. The error rate of the selection task decreased from 0.5 to 1.0s and reached the lowest point under the condition of 1.0s. Then, the error rate slightly rose up with a longer dwell time (1.2s). To further analyze error rates, we divided errors into false detections and missed detections. As the dwell time threshold increases, the false detection rate also decreases, in contrast, the missed detection rate increases. Based on the combined results above, overall, 0.8−1.0s is an optimal parameter range for the interface design. For the previous results based on eye-tracker, dwell time of 0.7-1s is considered sufficient (Majaranta, Aula, & Räihä, 2004), and the optimal range for both methods is quite compatible.

In most cases, the closer distance setting resulted in lower task completion times and error rates. Consistent with the objective evaluations, about half of the participants rated 45cm as the most comfortable distance, followed by 55cm. The vast majority (approx. 90%) of participants felt that the accuracy is higher at a distance of 45cm, compared to the other two distance conditions.

Although there were no significant differences in terms of the size of the central area in relation to the task completion time and error rate, the descriptive results show that the medium-sized central area condition achieved slightly shorter task completion time and lower error rate than that of the small- and large-sized central area conditions.

To apply the GaVe interface in real-world applications, future research should consider, first, the screen size. The display used in this study is relatively small. A larger screen size is expected to improve the correct detection rate. Second, head movement was not fully considered in our interface design, potentially limiting the real-world use of the interface, where head movements should be considered during gaze estimation to achieve a more robust interaction. In addition, individual height differences between users can also affect the usability of the system. This can be optimized by the automatic adaptation of camera height to a user's height to improve both face detection and gaze estimation, as well as user experience. This study focuses on a preliminary webcam-based gaze interaction design using a public vending machine as a user case, however, the whole system needs to be more refined, such as basket, payment. Last but not least, since eye movement data reveals implicit information of user, such as biometric identity, emotional state, interests etc., the privacy implications of eye tracking should be considered when using such method in public display (Kröger, Lutz, & Müller, 2019)

# Conclusion

In this paper, we conducted a proof-of-concept study for a hands-free input method based on gaze estimation using a webcam. GaVe interface was designed based on dwell time using this proposed method. Users can easily interact with the gaze-based interface after a 2s one-point calibration. As a touchless control modality, this interface design can improve the hygiene of using public displays, especially during the COVID-19 pandemic.

Based on the results of the user study, we draw the following conclusions for the design of public gaze-based interfaces: (1) A moderate distance needs to be considered. In this experiment, a user to interface distance between 45cm and 55cm is preferred and supports more robust detection, (2) the dwell time threshold could be set to $0.8-1.0s$, and (3) the size of the central area of the interface could be chosen as medium size, i.e., $13.4°×10.32°$ (at 45cm). In addition, our research can provide guidance on structuring the interface design for touchless ordering services in similar applications, such as ticket vending machines, automatic coffee machines, and parking meters, as the number of selectable items can be decreased by inserting additional selection rounds.

## Ethics and Conflict of Interest

The author(s) declare(s) that the contents of the article are in agreement with the ethics described in http://biblio.unibe.ch/portale/elibrary/BOP/jemr/ethics.html and that there is no conflict of interest regarding the publication of this paper.

## Acknowledgements

The publication of this article was funded by the Open Access Fund of Leibniz Universität Hannover.